\documentclass[floatfix]{emulateapj}
\usepackage{epsfig}

\begin{document}

\title{Structure and evolution of Zel'dovich pancakes as 
				probes of dark energy models}

\author{P.~M.~Sutter} \email{psutter2@uiuc.edu}
\affil{Department of Physics,
	     University of Illinois at Urbana-Champaign,
            Urbana, IL 61801-3080}

\and

\author{P.~M.~Ricker} \email{pmricker@uiuc.edu}
\affil{Department of Astronomy,
       University of Illinois at Urbana-Champaign,
             Urbana, IL 61801\\
		National Center for Supercomputing Applications,
      University of Illinois at Urbana-Champaign,
            Urbana, IL 61801}

\begin{abstract}  
We examine how coupled dark matter and dark energy modify the 
development of Zel'dovich pancakes.
We study how the various effects of these theories, 
such as a fifth force in the dark sector and a modified 
particle Hubble drag, produce variations in the 
redshifts of caustic formation and the present-day 
density profiles of pancakes.
We compare our results in direct simulation to 
a perturbation theory approach for the dark energy 
scalar field.
We determine the range of initial 
scalar field amplitudes 
for which perturbation theory is accurate 
in describing the development of the pancakes.
Notably, we find that perturbative methods which neglect 
kinetic terms in the scalar field equation of motion 
are not valid for arbitrarily small perturbations. 
We also examine whether models that 
have been tuned to match the constraints of 
current observations can produce new observable effects
in the nonlinear structure of pancakes.
Our results suggest that a fully realistic three-dimensional 
simulation will produce significant new observable features, 
such as modifications to the mass function and halo 
radial density profile shapes, that can be used to 
distinguish these models from standard concordance cosmology and from each other.
\end{abstract}

\maketitle

\section{Introduction}
\label{sec:introduction}
Dark energy is perhaps the most profound and essential 
mystery in modern cosmology. 
While the $\Lambda$CDM 
cosmological model has proven very successful in 
explaining and predicting many features of our universe, 
such as the 
fluctuations in the cosmic microwave background~\citep[eg.][]{deBernardis}, 
the large-scale matter distribution~\citep[eg.][]{Percival}, 
and 
distance measurements to type Ia supernovae~\citep{Perlmutter,Riess}, the physics 
of the dark sector (dark matter and dark energy), which comprises roughly ninety-six 
percent 
of the energy density of the universe, is 
largely unknown. 

 Currently, there are too few 
observational constraints to determine the precise 
nature of the dark energy 
and any possible interactions it might have with dark and baryonic 
matter~\citep{Bean}.
However, we can use 
simulations of nonlinear structure 
formation to explore the consequences of plausible
dark energy models, 
 including those that propose 
couplings between dark matter (DM) and dark energy (DE)~
\citep[see][and references therein]{Alcaniz}. 
Models that have particle physics motivations 
often predict such couplings~\citep{Amendola2}.
Such theories are 
intriguing because they might provide a resolution to current 
cosmological problems, 
such as the so-called coincidence problem~\citep{Zimdahl,Amendola3}, 
and the observed emptiness of the voids~\citep{Farrar}, 
the latter of which was confirmed numerically by~\cite*{Nusser}.

An important goal of coupled DM-DE simulations 
is to identify observational methods to test these theories. 
The observables studied to date include 
luminosity distances~\citep{Amendola}, 
the growth of matter perturbations
~\citep{Olivares,Koivisto}, the abundance of clusters~\citep{Manera}, 
and the Sandage-Loeb test~\citep{Corasaniti}.
Not only can we use such simulations to search for additional 
observable features, 
but we can also use them to find ways to distinguish models and determine 
the validity of perturbation methods. 

Much of the work to date has followed the framework 
established by~\cite*{Farrar}.
This model uses a 
dynamical scalar field to provide the dark energy, and it allows that field 
to couple to dark matter via a Yukawa interaction.
Although there are some 
issues with models of this type~\citep{Doran}, 
they provide useful foils 
for studying the possible effects that other 
models, which are complicated but more robust, 
might predict.
These models would include the model discussed 
by~\cite*{Huey} and the two-family model considered in Farrar and Peebles.  

Some of the previous work has examined structure formation 
with modifications due to a fifth force, albeit at low 
spatial and force resolution~\citep{Nusser,Maccio}. 
However, we believe 
that an incremental approach that analyzes the various 
effects of DM-DE interactions on simpler initial conditions is crucial 
before tackling more realistic initial conditions. This approach 
allows the effects observed in more realistic simulations 
to be understood and generalized.
Therefore, in this paper, we will study the effects of 
coupling between dark matter and dark energy 
on the development of Zel'dovich pancakes. Zel'dovich pancakes are 
the well-known solutions to the problem of 
the gravitational collapse of
one-dimensional, sinusoidal, plane-wave, 
adiabatic density perturbations~\citep{Zel}. Since they
are well-studied in a variety of contexts 
not involving DM-DE interactions 
\citep[for example,][and others]{More,Gnedin,Yuan,Valinia,Anninos1,Anninos2}, 
we can more easily understand the effects of 
additional physics on structure formation, laying the 
groundwork for a more complete three-dimensional study.

In Section~\ref{sec:model} we discuss
the relevant equations, the effects we will study, 
and our numerical techniques.
Section~\ref{sec:effects} 
discusses the role that the exotic physics 
introduced by the Farrar and Peebles model, such as a DM particle fifth force and 
time-dependent DM particle mass, 
plays in structure formation.
We are careful to separate the various effects predicted by the 
model. Even if this particular example does not withstand 
closer scrutiny, unrelated theories may predict 
similar effects.

In addition, so far many researchers have exploited perturbation theory to 
treat fluctuations in the DE scalar field. However, we do not know 
\emph{a priori} the validity of this approach, especially as we begin to 
explore the nonlinear consequences of this theory. 
In Section~\ref{sec:validity}
we will compare the structure formation results 
from perturbation theory alone  
to the results from doing a complete nonlinear analysis. 
We will also discuss 
the regime where perturbation theory is most valid in 
accurately predicting structure.

Many of the theories of this type have adjustable parameters, and 
these parameters must be adjusted to match $\Lambda$CDM predictions, at 
the risk of being indistinguishable from it. There are also 
usually several unique combinations of parameters that provide similar, 
if not identical, results.
In Section~\ref{sec:distinguish} we attempt to find ways 
to distinguish models that remain 
indistinct in perturbation theory. 
Also, we will determine if nonlinear effects can distinguish 
models from standard cosmology even when effects based on 
perturbative methods cannot.

\section{The Model}
\label{sec:model}
\subsection{Coupled Dark Matter and Dark Energy}
Following Farrar and Peebles, we will consider 
a dark energy (DE) scalar field $\phi$ with action
\begin{equation}
\label{eq:DEaction}
	S_{DE} = \int d^4 x \sqrt{-g} \left[ \frac{1}{2} 
		\phi_{,\nu} \phi ^ {,\nu} - V(\phi) \right],
\end{equation}
and a single classical nonrelativistic dark matter (DM) 
particle family with action for the $i$th particle
\begin{equation}
\label{eq:DMaction}
	{S_{DM}}_i = - \int y \left| 
		\phi\left( x_i \right) - \phi_\ast \right| 
		\sqrt{g_{\mu \nu} dx^{\mu}_{i} dx^{\nu}_{i}},
\end{equation}
where $y$ is the dimensionless Yukawa interaction 
strength and $\phi_\ast$ is a constant providing an 
intrinsic mass to the DM 
particle. We will set $\phi_\ast = 0$, so that 
the DM particle mass is due entirely to the field value.
In both of the above actions and throughout, we have set $\hbar = c = 1$, and 
we will ignore baryons. 

Assuming a spatially flat Friedmann universe, we can 
use equation~(\ref{eq:DMaction}) to
obtain a comoving DM particle 
equation of motion in one dimension:
\begin{equation}
\label{eq:dmEoM}
\dot v  + \left( 2 \frac{\dot{a}}{a} + \frac{\dot{\phi}}{\phi} \right)
 v = - \frac{\partial \Phi}{\partial x} - \frac{1}{a^2} \frac{1}{\phi} \frac{\partial \phi}{\partial x}  .
\end{equation}
Here, $\Phi$ is the normal comoving gravitational potential, 
$a$ is the scale factor, and $v$ is 
the comoving particle velocity, 
and $x$ is the comoving position. Throughout, dots 
refer to derivatives with respect to proper time $t$. 
The field action in equation (\ref{eq:DEaction}) gives the evolution of $\phi$:
\begin{equation}
\label{eq:phiEoM}
	\ddot{\phi} - \frac{1}{a^2} \nabla^2 \phi + 3 \frac{\dot{a}}{a} \dot{\phi} 
	  + \frac{d V}{d \phi} + \frac{\rho}{\phi}a^{-3}= 0,
\end{equation}
where $\rho$ is the DM particle comoving density. 

The comoving potential satisfies the Poisson equation:
\begin{equation}
\label{eq:poisson}
 \nabla^2 \Phi = \frac{4 \pi G}{a^3} \left( \rho - \overline{\rho} \right).
\end{equation}
Here and throughout, an overline indicates a spatial average.

The Friedmann equation, neglecting radiation, curvature, and 
baryonic terms, is now
\begin{equation}
\label{eq:friedmann}
	\left( \frac{\dot{a}}{a} \right)^2 = 
	H_0^2 \Omega_m \frac{\overline{\phi}}{\overline{\phi}_0} a^{-3} + 
	\frac{8 \pi G}{3} \left[ \frac{1}{2} \left( \frac{d \overline{\phi}}{d t} 
	\right)^2 + V(\overline{\phi}) \right].
\end{equation}
A subscript of $0$ denotes the present-day value.
The first term on the right-hand side reflects the contribution 
of the DM with its time-dependent mass. 
The terms in brackets are, respectively, the kinetic 
and potential energies of the DE scalar field.

We do not know the 
initial conditions of the field, but we do know that today the field 
behaves as a cosmological constant, so the potential term dominates and 
has a value 
\begin{equation}
	V(\overline{\phi}_0) = \Omega_\Lambda \rho_{crit}. 
\label{eq:potentialToday}
\end{equation}
Also, at early 
enough times, Farrar and Peebles found that 
equation (\ref{eq:phiEoM}) reduces to
\begin{equation}
\label{eq:initialPhiDot}
	\frac{d \phi} {d t} = - \frac{H_0^2}{G} 
	\frac{3 \Omega_m }{8 \pi \phi_0} \frac{1}{a^3} t,
\end{equation}
which we use to set the initial condition for $\dot \phi$. 

We can identify four unique ways in which the 
extra interactions modify structure formation.
First, the DM particle mass directly depends on 
the field value, so we may write 
the ratio of the modified mass to its present-day value as
\begin{equation}
\label{eq:mass}
 \eta \equiv \frac{m_{DM}}{m_{DM,0}} = \frac{\phi}{\phi_0}
\end{equation}
Secondly, the interactions modify the Hubble drag found 
in the DM equation of motion, so that its 
ratio to the drag in standard cosmology is
\begin{equation}
\label{eq:drag}
 \gamma \equiv \frac{2 \dot{a}/a + \dot{\phi}/ \phi}{2\dot{a}/a}.
\end{equation}
Next, we notice a modified particle acceleration due 
to a fifth force in equation (\ref{eq:dmEoM}). 
We will define 
\begin{equation}
\label{eq:force}
 \beta \equiv \frac{\frac{\partial \Phi}{\partial x}  + 
 								\frac{1}{a^2} \frac{1}{\phi} \frac{\partial \phi}{\partial x} }
 									{\frac{\partial \Phi}{\partial x} }.
\end{equation}
Finally, the dynamic scalar field itself 
indirectly affects structure formation 
via a time-varying $\Omega_\Lambda$ in the Friedmann equation. We define
\begin{equation}
\label{eq:field}
 \delta \equiv \frac{
    \frac{1}{2} \left( \frac{d \overline{\phi}}{d t} 
	\right)^2 + V(\overline{\phi})
 }{V \left( \overline{\phi}_0\right)}.
\end{equation}

If the fluctuations in the field are small enough, we may do 
perturbation theory. Farrar and Peebles found that in this regime, 
we may replace $\phi(x)$ with 
a single spatial average, $\phi_b$ and a 
sufficiently small perturbation field $\phi_1(x)$. 
We may then write equation~(\ref{eq:phiEoM}) as
\begin{equation}
\label{eq:phibEoM}
	\ddot{\phi_b} + 3 \frac{\dot{a}}{a} \dot{\phi_b} 
	  + \frac{d V}{d \phi_b} + \frac{3 \Omega_m H_0^2}
		{8 \pi G \phi_{b,0}}a^{-3}= 0.
\end{equation}
Also, the fifth force ratio appears instead as
\begin{equation}
\label{eq:fifthPert}
	\beta_{pert} \equiv 1 + \frac{1}{4 \pi G \phi_b^2}.
\end{equation}

We will contrast our results with standard concordance cosmology, which we 
achieve by setting $\phi$ to a single value satisfying 
equation (\ref{eq:potentialToday}) and by  
preventing $\phi$ from evolving dynamically.

We have the freedom to choose an appropriate potential $V(\phi)$. 
Although there are many potentials in the literature, such as the 
exponential~\citep{RatraAndPeebles}, power-law~\citep{PeeblesAndRatra}, 
and power-law and sine~\citep{Dodelson},
we will adopt the power-law potential found in Farrar and Peebles:
\begin{equation}
\label{eq:powerLaw}
	V(\phi) = K / {\phi}^{\alpha},
\end{equation}
where we are also free to choose the constants $K$ and $\alpha$.
Potentials like this lead to reasonable behavior, i.e. 
potential-dominated solutions at the present epoch. 
Again, we chose this potential merely as an example.

\subsection{Numerical Techniques}
For our simulations we developed a one-dimensional 
$N$-body particle-mesh code. 
We used cloud-in-cell mapping for interpolating between 
the mesh and particles~\citep{Hockney}, and 
a second order leapfrog integration scheme for particle advancement. 
In one dimension and with finite differencing, 
we can solve the Poisson equation exactly using a 
Thomas algorithm~\citep{Thomas}, modified for 
periodic boundaries via the Sherman-Morrison formula~\citep{Sherman}.
We discuss the details of our scheme for solving the scalar field 
and scale factor in the appendix.

For all calculations, 
we used $\Omega_m = 0.26$, $\Omega_\Lambda=0.74$, 
and $H_0 = 100 \mbox{ }h = 71 \mbox{ km s}^{-1} 
\mbox{ Mpc}^{-1}$. All runs took place in a 
one-dimensional box of length $10 \mbox{ } h^{-1} \mbox{ Mpc}$ per
side, with $65,536$ particles to represent 
the dark matter and $8,192$ zones 
for the Poisson solver and the scalar field.
Since we used finite-differencing when forming the gradient 
to construct the fifth force, we required a 
large particle-zone ratio to dampen noise 
in the density field, which couples to the scalar field.

All simulations used the same initial conditions. 
We distributed particles evenly throughout the grid (described by 
position $q_i \equiv i \Delta x$) and perturbed them 
using the Zel'dovich approximation:
\begin{eqnarray}
	x & = & q + \frac{2}{5} a A \sin{(k q)} \\
	v & = & \frac{2}{5} \dot{a} A \sin{(k q)},
\end{eqnarray}
where $k=2\pi / \lambda$ is the comoving wavenumber 
of the perturbation.
The amplitude $A$ can be written in terms of the redshift 
$z_c$ of the formation 
of the first caustic: 
\begin{equation}
	A = -5(1+z_c)/(2k).
\end{equation}
We chose the initial perturbation amplitude 
such that the first caustic would 
form at $z_c=5.0$ for an $\Omega_m = 1.0$ universe, and 
we chose the comoving wavelength 
of perturbations to be $\lambda = 10 \mbox{ }h^{-1} \mbox{ Mpc}$.
These choices are arbitrary, but commonly used in the literature. 
All computations start 
at a redshift of $z=50$ with DM particle masses 
determined by equation~(\ref{eq:mass}).

We do not know the initial conditions of $\phi$ in advance, 
so we must guess initial conditions 
and iterate until we meet the condition $\dot{a} = H_0$ 
at the present epoch.
Based on the comments made by Farrar and Peebles, we chose four 
combinations of the potential parameters $K$ and $\alpha$ 
that yield reasonable behavior. 
Table~\ref{tab:FandP} lists the parameters, the guessed 
initial field value at $z=50$, and the field value today as calculated 
from equation~(\ref{eq:phiEoM}). We performed these calculations in 
perturbation theory. 
We selected these parameters for behaviors that were 
consistent with current observations, but provided 
unique evolutions of the scalar field for 
comparison.
 	         
\begin{table}
  \centering
	\begin{tabular}{|c|c|c|c|c|} 
		\hline
	Label & $ \alpha $ & $K (G^{1+\alpha/2} / H_0^2)$ 
	      & $\phi_{\mbox{init}} (G^{1/2})$ & $\phi_0 (G^{1/2})$\\
	\hline
	A & $-2$ & $0.03$   & $1.89$   & $1.72$  \\
	B & $-2$ & $0.0057$ & $4.02$   & $3.94$  \\
	C & $6$  & $2.0$    & $1.80$  & $1.68$  \\
	D & $6$  & $280.0$  & $3.89$   & $3.83$  \\
	\hline
	\end{tabular}
	\caption{Simulation parameter choices.}
	\label{tab:FandP}
\end{table}
 
\section{Analysis of Effects in Perturbation Theory}
\label{sec:effects}
We now illustrate the effects of these 
additional interactions on structure formation. 
To highlight each individual consequence of the DM-DE interaction, 
we separately analyze the resulting structure when including the 
effects of the variable mass, 
the modified Hubble drag, the fifth force, and the 
dynamical field.
Finally, we will examine the final structure when 
these effects are combined.
In all that follows, we highlight the behavior 
when using the potential 
parameters $K = 0.03$ and $\alpha = -2$ 
(parameter set $A$) as an example.

\begin{figure}
	\plotone{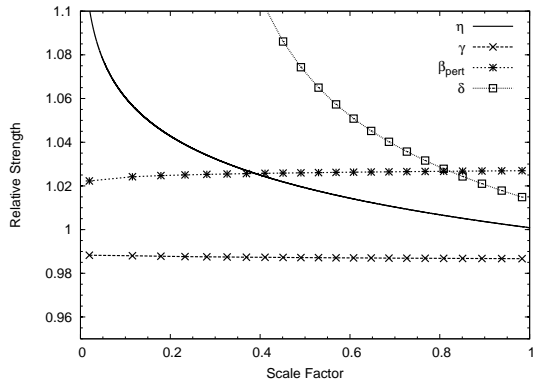}
	\caption{Comparison of additional effects due to the 
		DM-DE interactions. Shown are relative values defined 
	        above as $\eta$ (solid line), $\gamma$ (line with hatches), 
		$\beta_{pert}$ (line with stars), 
	        and $\delta$ (line with boxes). 
	        The symbols are only to aid in distinguishing the lines.}
\label{fig:compareEffects}
\end{figure}

Figure~\ref{fig:compareEffects} shows the 
evolution of $\eta$, $\gamma$, $\beta_{pert}$, 
and $\delta$ as
 functions of scale factor for parameter set $A$. 
On average, the fifth force is 
about three percent of the gravitational force. 
We expect the fifth force to play a significant role 
throughout cosmic history. 
The Hubble drag is reduced by about one percent. The particle 
mass starts at ten percent greater than its present-day value, 
but by a scale factor of $0.5$ it is only 
two percent larger. We expect the modified mass to contribute to 
structure mostly in the early universe, 
but by the present day its effects will not be significant. 
In the early universe, the dynamic 
field gives a large $\Omega_\Lambda$, and hence a 
more rapid expansion, which should delay 
caustic formation relative to standard cosmology. 
Notice that our calculations do not give 
precisely the required values of $\Omega_\Lambda$, 
but are less than two percent off. 
We fixed our field at late times 
such that the energy density of the field was purely 
in the potential, when in reality 
the kinetic term, while very small, 
is not entirely negligible. Fortunately, this small 
error does not in any way significantly alter structure formation.
  
\begin{figure}
	\plotone{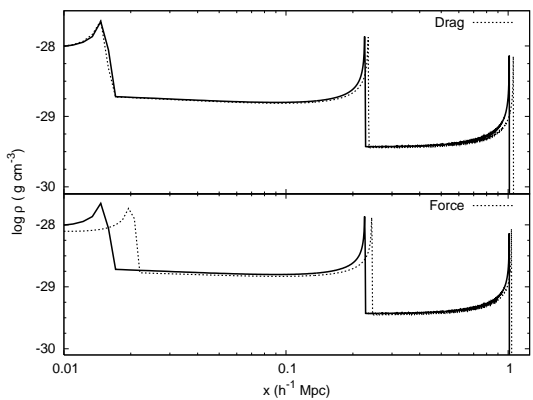}
	\caption{Dark matter density profile at $z=0$. 
		The solid line in all plots is the standard 
		cosmology. The dotted line indicates the 
		effects of the additional interactions: 
		the top panel shows only the result with 
		the drag term included, and the bottom panel
		shows only the fifth force included.
		Here, $x$ is the distance from the midplane 
		of each pancake.
		The numerical shot noise in the density near $x=1$ 
		is negligible and unimportant.}
\label{fig:compareDensityProfile}
\end{figure}

We show the density profile at $z=0$ in 
Figure~\ref{fig:compareDensityProfile}, modified 
by the individual additional effects of the DM-DE interaction.
We do not show the result from an evolving mass and 
dynamical field, because these profiles are 
hard to distinguish from standard cosmology. 
When isolating the individual effects 
of the drag, fifth force, and dynamic field, 
we forced the scalar field to evolve as if the scale factor 
obeyed the modified 
Friedmann equation in equation~(\ref{eq:friedmann}); however, the actual scale 
factor evolved according to concordance cosmology in these comparisons. 
We did this because without the dynamic mass in 
the Friedmann equation, the 
scalar field would not reach its required value today.
 
We can easily understand the modifications due to the 
fifth force: an additional attractive 
force causes the caustics to form earlier, 
and hence each peak in the density profile at $z=0$ 
is farther away from the center. The effect on each peak is similar:
a fifth force only in the early universe will 
affect mostly the first two caustics, 
but since the force ratio remains roughly constant 
even in late times, the latest peak also shifts.
The overall size of the pancake is then larger. 

Due to the modified drag term alone, the first two caustics form earlier, 
but the latest caustic forms at the same time as in 
standard cosmology.
This requires more explanation.
The Hubble drag term is most important in the early universe. 
Due to the reduced drag, particles 
move faster, and reach a greater turnaround radius. 
So the first two caustics form earlier for the 
modified cosmology. However, at later times, 
the drag is no longer significant, and 
the third caustic forms at nearly the same 
redshift as in standard cosmology.
Also, it appears that modifications to the drag 
are very important: even small deviations in the 
redshift of caustic formation lead to significant 
difference in the final peak locations.

The increased mass creates tension between 
two opposing influences on structure formation. 
Since the particles already in halos are more massive, 
they create a deeper potential 
well and encourage other particles to fall in faster, 
thereby causing caustics to form 
earlier. On the other hand, a larger $\Omega_m$ in the 
Friedmann equation accelerates 
the expansion in the early universe, which increases 
the Hubble drag and dampens 
structure formation. 
We found that the expansion effect dominates, but 
not significantly, and caustics form only slightly 
later than in standard cosmology.
Also, since the DM particle is more massive in the past, 
the regions near early-forming caustics have a higher density, 
but by $z=0$ the mass is the same as in standard cosmology, so the overall 
amplitude of the density profile does not change. 

We must contrast this evolving-mass 
universe with a $\Lambda$CDM cosmology having simply a larger $\Omega_m$. 
In such a universe, the competing 
effects of increased mass roughly cancel each other out, 
and caustics form at 
similar redshifts to those in standard $\Lambda$CDM. 
In an evolving-mass universe, 
however, as the mass decreases, 
the halo potential wells 
get shallower, but the universe has already expanded 
more, so outlying 
particles take longer to reach the halo, 
and caustics correspondingly form later. 
At lower redshifts, 
the particle masses are nearly their present-day 
values, so this effect is not as significant, and the final 
peak locations are slightly closer to the pancake midplane.

Finally, the dynamic field has the expected result: 
more rapid early-universe expansion 
pulls particles away from each other, and caustics take longer to form.
As expected, this effect is nearly negligible 
in the late universe, as the field 
approaches its present-day value. 
Unlike the differences between an evolving-mass 
universe and a constant-mass universe, 
the structure modifications from a dynamic 
field are, in this case, nearly indistinguishable from a universe with 
a roughly ten percent larger, but still constant, $\Omega_\Lambda$.

\begin{figure}
	\plotone{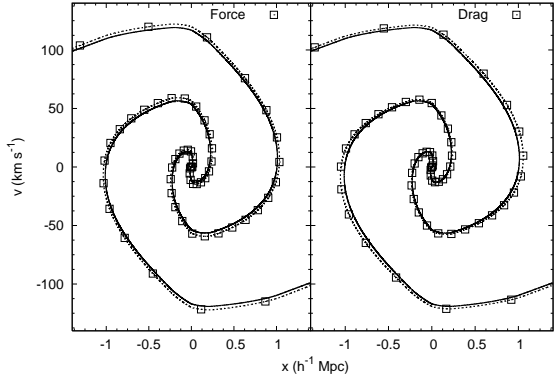}
	\caption{Dark matter particle phase plots at $z=0$.
		The solid line in all plots is from concordance cosmology, 
		and the dotted lines with open squares 
		are the results from including the 
		effects indicated in each plot. 
		The open squares only represent select data points, to aid 
		in distinguishing the lines.
		We have omitted the plots 
		with dynamic field and dynamic mass effects, since these 
		have no discernible influence.}
\label{fig:comparePhase}
\end{figure}

We can also examine the effects on particle 
velocities due to these additional interactions. 
Figure~\ref{fig:comparePhase} is a DM particle phase plot at $z=0$ with 
modifications due to the reduced Hubble drag and the fifth force, 
compared to standard cosmology. 
The dynamic field and variable mass terms have no effect, but
as expected, the reduced drag and fifth force increase particle velocities.
 
\begin{figure}
	\plotone{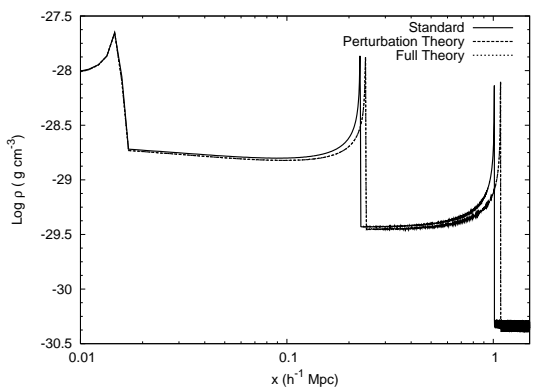}
	\caption{Dark matter density profile at $z=0$.
		The solid line is from standard cosmology. The dashed line is 
		the density from using perturbation theory, and the dotted line 
		is from using full theory. 
		This is including all effects.
		Note that perturbation theory is nearly 
		indistinct from the full theory.
		The numerical shot noise in the density near $x=1$ 
		is negligible and unimportant.}
\label{fig:compareStandardPertFull}
\end{figure}

When we combine results, in Figure~\ref{fig:compareStandardPertFull}, 
we see that the effects of the modified mass 
and the dynamic field tend to cancel out the fifth force. 
So, the final density profile most closely resembles the 
results from the drag term alone. 
The innermost caustic is the same distance from the midplane
of the pancake, 
but the outermost and middle peaks are roughly 
$15\%$ farther away.
The slope of the density profile 
remains undisturbed and the total number of peaks formed by the 
present epoch are the same.
 
\section{The Validity of Perturbation Theory}
\label{sec:validity}
So far, we have only considered perturbation theory, 
relying on a single value to describe the DE scalar field at 
a specific time. When we relax this constraint,  
each particle feels its own force, has its own drag, and has 
an independent mass. When solving the Friedmann equation, 
we must take a spatial average of the field, rather 
than using a single background value. Also, the expression for 
the fifth force changes 
to be directly proportional to gradients in the scalar field.

There is some ambiguity when considering the initialization 
of the scalar field. While 
the average value must match the perturbation theory background 
value, we are left with 
few clues to the initial wavelength and amplitude of the 
perturbations. So, using adiabatic reasoning, 
we initialized $\phi$ to have 
the same spectrum as the density field, which in our case 
is a single perturbation mode with 
wavelength $\lambda = 10 h^{-1} \mbox{ Mpc}$ at $z=0$. 
However, we freely chose the amplitude, and 
we found that to make an attractive fifth force,
which is required for a Yukawa-type interaction, 
we needed $\phi$ to have a phase opposite to that of the density field.

We must take care in the initialization of both the background 
value and amplitude of perturbations of the field when 
considering the full theory. First, the addition of the 
gradient term in the equation of motion for the field will 
affect the field's background evolution, and may force us to modify 
the initial value, so that $\phi$ still satisfies the 
present-day condition 
in equation~(\ref{eq:potentialToday}).
Fortunately, the gradients are not large, and 
this term does not modify the evolution of $\phi$ greatly. 
Hence, we constructed the field 
so its average value matched the background perturbation value.
It appears that perturbation theory is perfectly appropriate for probing the linear 
results of this model, such as estimates of the 
Hubble time and the location of peaks in the CMB power spectrum, 
which Farrar and Peebles discuss. 

\begin{figure}
	\plotone{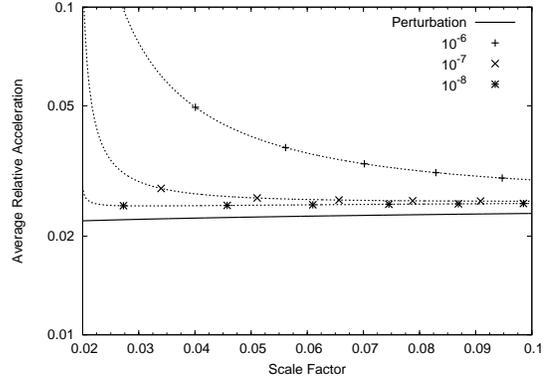}
	\caption{Average ratio of fifth force to gravitational 
		force ($\langle \overline{\beta} \rangle-1$)
		for perturbation theory (solid line), 
		compared to the full result with 
		varying initial amplitudes 
		(dotted lines with various marks, labeled in figure).
		The marks on the lines only represent select data points, to aid 
		in distinguishing the lines. 
		Shown are the average strengths up to the 
		given scale factor. }
\label{fig:compareFifth}
\end{figure}

Secondly, by varying the initial amplitude, we can change the 
strength of the fifth force at high 
redshift.~\cite*{Kesden,Kesden2} have shown that a constraint 
on any dark matter fifth force to within only a few percent 
of the gravitational force is observationally possible. 
The fifth force under perturbation theory agrees with this 
constraint: $\beta_{pert} \approx 1.03$ throughout cosmic history. 
Figure~\ref{fig:compareFifth} shows the space-averaged 
and time-averaged value of 
$\beta_{pert}-1$ and $\beta-1$, which 
we denote as $\langle \overline{\beta} \rangle - 1$, versus redshift for varying 
initial amplitudes. This is the average fraction of 
gravitational acceleration experienced by the DM particles due to the 
fifth force up to the given scale factor.
The field fluctuates rapidly in the early universe, and by 
time-averaging we can better examine the general 
force trends. 
The fluctuations in the field decay rapidly and 
asymptotically approach a constant value by redshift 
$z=45$. 
 
No matter the initial amplitude, the average 
perturbation theory fifth force ratio
agrees to within ten percent with the full result by the present epoch. 
However, the strength of the fifth force matters much 
more at high redshifts: different forces here can 
greatly affect structure formation. We found that an initial amplitude 
contrast of $\delta_\phi(z=50)=10^{-8}$ provides 
roughly the same order of magnitude additional 
acceleration as perturbation theory 
predicts at early times. 

For amplitudes smaller than $10^{-8}$, the results 
for the fifth force do not change. Since the scalar field 
couples directly to the dark matter density, the field 
will always contain perturbations. If we initialize 
the amplitudes below the threshold of $10^{-8}$, 
the density coupling will produce fluctuations in the 
scalar field spectrum, creating a fifth force.
However, larger amplitudes 
provide too much of an acceleration at early times. 
For example, a contrast of $\delta_\phi(z=50)=10^{-6}$ 
provides an average fifth force that is almost half as strong as 
gravity at early times.  
We found that if the initial contrast is larger than 
can be handled adequately by 
perturbation theory, then we get radically different structure formation, 
such as a very early first caustic and the development of four 
caustics by the present epoch. 
Note that these results are largely redshift-independent: 
the DM coupling will always end up dominating the spectrum 
of scalar field fluctuations after a short period of time. 
The large fifth force for amplitudes above $10^{-8}$ 
is due to the large artificial initial scalar field amplitude.
 
However, if the amplitude of 
perturbations is small in the early universe, 
then perturbation theory under-emphasizes the 
fifth force at high redshift.  
The fact that perturbation theory 
cannot describe large fluctuations makes sense.
However, we would expect that perturbation 
theory should accurately capture the behavior 
of all initial amplitudes below a certain value. 


With the above considerations, we examined the case of an initial contrast of $\delta_\phi(z=50)=10^{-8}$,
since this provides a fifth force large enough to affect structure, but 
which is not ruled out by observations. 
We compare perturbation theory 
to the results from a full dynamical field. 
We found that for this amplitude, perturbation theory 
is excellent in capturing the structure 
and evolution of halos. By construction, the average value 
of the dynamical field is the same as the background field 
in perturbation theory. Hence, the modified drag term and 
modified mass terms are almost identical.
Also, 
perturbation theory appears to be adequate for 
describing the evolution of the scalar field, 
so the effects of the dynamic field itself agree in the full theory.
Additionally, since fluctuations dampen with time, 
we only need to satisfy perturbation theory 
constraints in the early universe. We see in Figure~\ref{fig:compareStandardPertFull} 
that, for this particular amplitude of fluctuations, perturbation theory 
is excellent in describing the overall structure of the pancakes.

Overall, as Figure~\ref{fig:compareStandardPertFull} showed, 
perturbation theory seems to be adequate in describing
the larger features of the pancakes: 
the number of caustics and the location of all three caustics at $z=0$. 
With the notable exception of the fifth force, 
as we decrease initial scalar field amplitudes, 
perturbation theory becomes more accurate 
in providing an accurate solution. Even though the perturbation 
theory fifth force disagrees with the full theory by several 
percent, this is not enough to affect structure formation 
at our resolution.

\begin{figure}
	\plotone{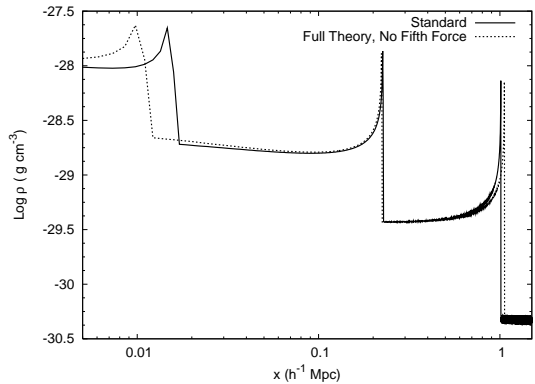}
	\caption{Dark matter density profile at $z=0$. 
		The solid line is the standard $\Lambda$CDM  
		cosmology, and the dashed line is the final result in full 
		theory when the fifth force is negligible.
		The numerical shot noise in the density near $x=1$ 
		is negligible and unimportant.}
\label{fig:compareStandardNoForce}
\end{figure}

It is interesting to note that we can produce significant 
changes in structure formation even with a negligible fifth force, 
such as might happen with a sufficiently screened potential. 
Figure~\ref{fig:compareStandardNoForce} shows the results 
from the full theory, but with no fifth force. In this case, 
we get stronger deviations in the innermost pancake substructure. 
Also, the second caustic forms slightly later than in 
standard cosmology, while the largest 
caustic is no different than when the fifth force is large.

\section{Distinguishability of Models} 
\label{sec:distinguish} 
The models chosen in Table~\ref{tab:FandP} are 
constructed to be consistent with current observations. 
For example, all these 
models predict values for the CMB peak locations that are 
within observational constraints. 

\begin{figure}
	\plotone{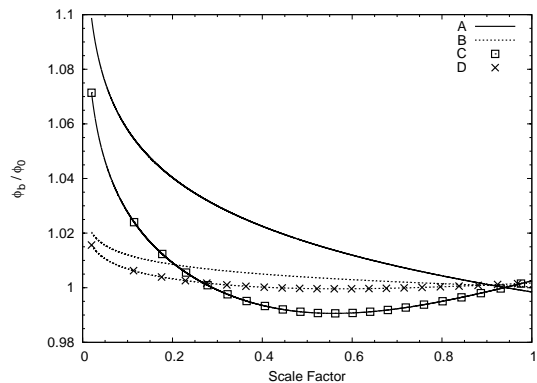}
	\caption{Evolution of $\phi_b$, relative to the 
	  present-day value, for the various 
		potential parameters labeled in 
		Table~\ref{tab:FandP}. 
		Note that the symbols are only to aid in distinguishing the lines.
		These 
		calculations are from perturbation theory.}
\label{fig:comparePhi}
\end{figure}

Figure~\ref{fig:comparePhi} 
shows how each of our chosen potential parameters in Table~\ref{tab:FandP} 
affects the evolution of $\phi$. 
These results agree with the perturbation results from Farrar and Peebles.
There are in general two classes 
of viable $\phi$ evolution tracks.
We see from Table~\ref{tab:FandP} that models such as  
$A$ and $C$ have a smaller initial value, but 
drop by roughly ten percent by the current epoch.
On the other hand, models such as $B$ and $D$ have a high 
initial value and do not change much throughout cosmic history.

\begin{figure}
	\plotone{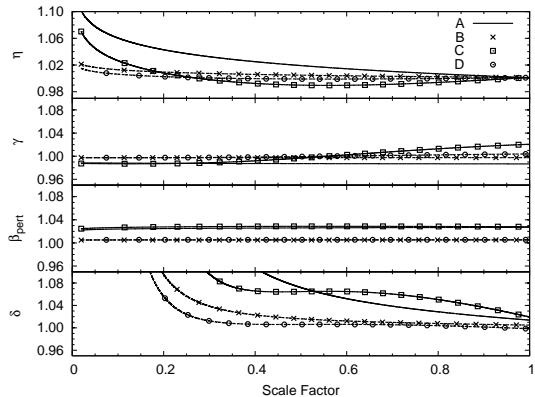}
	\caption{Evolution of 
		mass ($\eta$), 
		Hubble drag ($\gamma$), 
		fifth-force acceleration ($\beta_{pert}$), 
		and dynamical field ($\delta$) 
		for the various potential parameters labeled in 
		Table~\ref{tab:FandP}. 
		The symbols are only to aid in distinguishing the lines.
		These 
		calculations are from perturbation theory.}
\label{fig:comparePhiEffects}
\end{figure}

These two classes of parameter sets have interesting 
consequences for the behaviors of 
the DM particle mass, the modified Hubble drag, 
the fifth force, and the dynamical field itself. 
Figure~\ref{fig:comparePhiEffects} shows how 
the mass ratio ($\eta$) , 
the modified Hubble drag ($\gamma$), 
the fifth-force acceleration ($\beta_{pert}$), and 
the field energy density ratio ($\delta$) 
vary across the models in perturbation theory. For models of the second type, 
in which $\phi$ does not change much, the mass at $z=50$ is 
much closer to its present-day value, and hence this effect
on structure formation will be smaller than in models 
where $\phi$ varies strongly. Also, if $\phi$ does not 
vary much, then the Hubble drag will not be modified greatly.
Together, this makes sense: a set of parameters 
that favors a slightly-varying scalar field 
will look more like standard $\Lambda$CDM than those which do not.
Finally, for larger values of $\phi$, the fifth 
force is much weaker, since the 
force is proportional to $\phi^{-2}$ in perturbation theory.

\begin{figure}
	\plotone{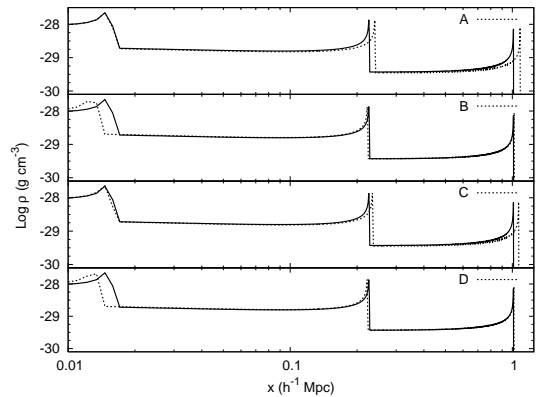}
	\caption{Dark matter density profile at $z=0$. 
		The solid line in all plots is the standard 
		cosmology. The dashed line indicates 
		the result in full theory, 
		for the various potential parameters labeled in 
		Table~\ref{tab:FandP}.}
\label{fig:comparePhiDensityProfileFull}
\end{figure}

These behaviors play out accordingly in the final structure at $z=0$, as 
Figure~\ref{fig:comparePhiDensityProfileFull} shows. 
These calculations were done using the full theory. 
When performing the analysis in the full dynamical theory, 
we again have to 
be careful when setting the initial scalar field amplitude. 
For all cases, 
we set an initial contrast of 
$\delta_\phi(z=50)=10^{-8}$.

Models $B$ and $D$ do not differ 
much from standard cosmology or each other. Models $A$ and $C$, 
while both different from 
$\Lambda$CDM, remain mostly indistinguishable, 
except that parameter set $A$, whose 
changes in $\phi$ are most drastic, produces caustics 
that are slightly farther away from the pancake midplane. 
Even though $\phi$ 
is increasing at the present epoch in parameter set $C$, 
this change happens 
at late times, and so it does not have a significant 
effect on the resulting structure. 

We also performed this study in perturbation theory, and 
we found that accounting for the effects of the full 
dynamics do not increase the distinguishability of the models.


\begin{figure}
	\plotone{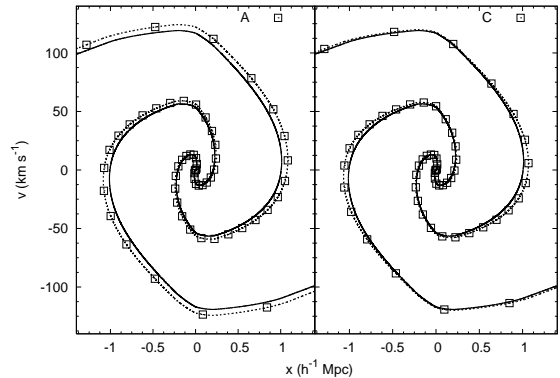}
	\caption{Dark matter particle phase plots at $z=0$.
		The solid line in all plots is from concordance cosmology, 
		and the dotted lines are the results from various 
		model parameters, which are labeled in the plots.
		These were calculated in the full theory.
		The symbols are only to aid in distinguishing the lines.}
\label{fig:comparePhaseModels}
\end{figure}

Surprisingly, we do notice some additional distinguishing features 
when examining the phase plots, as in Figure~\ref{fig:comparePhaseModels}. Since models $B$ and $D$ 
remain nearly identical to concordance cosmology, we do not display them. 
While models $A$ and $C$ share many common features, 
the drag term $\gamma$ 
increases at late times in model $C$. 
This eventually slows down particles at low redshifts and pushes 
the peak velocities closer to concordance cosmology.
Hence, the higher peak particle velocities 
distinguish model $A$, whereas simply considering 
pancake density profiles may not.

\section{Conclusions}
\label{sec:Conclusion}
These one-dimensional simulations have clearly demonstrated that 
models of interacting dark matter and dark energy affect the 
growth and structure of plane parallel perturbations. Various consequences of 
these theories play important roles at different stages in halo 
evolution. Larger fifth forces and modified Hubble drag terms 
in the high-redshift universe greatly alter early structure formation, 
while an evolving particle mass can affect even late-forming structures.   
Ordinary quintessence, which is a dynamic field alone, 
does not significantly alter the formation 
of pancakes. Of all the effects, a modification to the Hubble 
drag gives the largest deviations in the resulting structure.

We have found that models that are specifically tuned 
to match current observational constraints, such as the 
CMB peak locations and equation of state parameters, 
produce significant deviations from standard 
$\Lambda$CDM in the formation of pancakes.

We have also found that an approach  
based on perturbation theory is adequate for understanding the 
general evolution of the scalar field.
However, perturbation theory does not seem to be appropriate 
at high redshift. Even for arbitrarily small amplitudes, 
the fifth force in the full theory is larger than that obtained 
by perturbative techniques.
This behavior deviates from the general 
pattern of perturbation methods, in that 
it does not accurately describe all characteristics
for amplitudes smaller than a threshold value. 

This discrepancy arises from the fact that the 
fifth force directly depends on the gradient of the 
field. In the perturbative method described 
by Farrar and Peebles, the kinetic terms of the perturbation 
field equation of motion are dropped, producing a 
Poisson equation for the perturbation field and leading to the 
simplification of the fifth force expression found in equation~(\ref{eq:fifthPert}). 
This simplification is valid at low redshifts, but at high 
redshifts the kinetic terms are still important.
Thus, any realistic spectrum of perturbations 
for the scalar field which 
ignores the kinetic terms may be invalid for the 
very largest-scale perturbations.
It also appears that the initial scalar field 
amplitude does not serve as a suitable ``small'' 
parameter for governing the appropriateness 
of perturbation theory.

Because we can freely choose the initial amplitude, 
full theory gives us more freedom to study the effects 
of various strengths of a fifth force. 
In the context of this model, 
this can also work in reverse: 
working within these frameworks,
constraints on a fifth force in the dark sector could 
lead to limitations on the amplitudes 
of a DE scalar field in the early universe.

Anninos and Norman have extensively studied the effects on baryons in the formation of 
pancakes~\citep{Anninos1,Anninos2}. 
While baryons appear to modify only the last-forming caustic, DM-DE 
interactions affect all caustics. 
Hence, the modifications to the last caustic may be masked by the baryon dynamics. 
Also, 
baryons do not appear to significantly modify dark matter particle velocities, while 
some of the models considered above produce significant variations. 

The structural differences described above are relatively small. 
However, in three dimensions these effects should appear as 
percent-level differences in the statistical properties of large samples of dark 
matter halos. The various effects could produce different halo mass 
functions, which could be compared to other 
high-resolution mass functions, 
such as the $\Lambda$CDM ones considered by~\citet{Warren}.  
In fact,~\cite*{Mainini} have shown that 
DM-DE interactions do modify analytic mass functions, 
and a direct simulation could be compared against 
these results. DM-DE 
interactions may also affect halo substructure, 
in which case they would impact 
NFW profiles~\citep{NFW} and distributions of 
concentrations in halo catalogs~\citep{Lukic, Reed}.

The present work focused on a single model and a 
single potential, simply to analyze the 
feasibility of a direct simulation approach.
With a full three-dimensional simulation, we can 
also examine the results of using other potentials, or 
even more complicated models.
With future simulations, we 
can also contrast these structure formation results to the 
\emph{N}-body results from other theories of DE, 
such as modified General Relativity, which 
have been examined in~\citet{Stabenau}.

For the models with the largest deviations from $\Lambda$CDM, we notice 
that the sign of the difference of the caustic peak location changes.
This implies that the shape and evolution of the mass function for these
cosmologies will change relative to $\Lambda$CDM, 
not simply the amplitude. This holds the most hope for 
observations, particularly for surveys such as the DES~\citep{Annis}.
Most notably, we found that we will still get
significant modifications to pancake structure even if the fifth force is negligible. 
Indeed, the modifications to the mass function may even 
be more significant without a fifth force.
Since many recent efforts have concentrated mostly on 
constraining the fifth force~\citep{Kesden,Farrar2,Bertolami,Guo}, this result clearly shows 
that we should not ignore the other 
consequences, such as modifications to the 
Hubble drag, of these models.
Also, the phase plots revealed another observable effect: the large-scale matter velocity field, 
which is also accessible to the DES survey. 
The feasibility of constraining these models with this behavior could be tested with 
velocity correlations in full three-dimensional simulations.

However, there are several computational and 
theoretical challenges in developing 
and analyzing full 
three dimensional simulations. For example, there should be a stronger 
theoretical understanding of the initial conditions of the field. 
While implicitly solving the scalar field is 
relatively straightforward in one dimension,
requiring a simple Thomas algorithm, 
a full three dimensional solver would be much more complex, 
probably requiring a multigrid solver that could easily be 
combined with existing parallel adaptive-mesh codes
such as FLASH~\citep{Fryxell} or GADGET-2~\citep{Springel}. Since there 
might be interesting changes in halo substructure, simulations 
will require very high resolution. 
Also, simulations will require many halos 
to get significant statistical results.

Fortunately, none of these issues are intractable, and this 
approach holds much hope for providing a strong, consistent 
method of analyzing the nonlinear effects of these myriad 
dark energy models.

\acknowledgements
The authors would like to thank Luke Olson, Ben Wandelt, 
and Greg Huey for enlightening and valuable discussions.

The authors acknowledge support under a Presidential Early 
Career Award from the U.S. Department of Energy, 
Lawrence Livermore National Laboratory (contract B532720).
Additional support was provided by a DOE 
Computational Science Graduate Fellowship 
(DE-FG02-97ER25308) and the National Center for 
Supercomputing Applications.

\appendix
\section{Numerically Solving the Scalar Field Equation}
\label{app:numerics}
We have
\begin{eqnarray}
\label{eq:phi}
\ddot{\phi} - \frac{c^2}{a^2} \nabla^2 \phi + 3 \frac{\dot{a}}{a} \dot{\phi} + \frac{d V}{d \phi} 
 + \frac{\rho}{\phi}a^{-3}& = & 0.  
\end{eqnarray}

We will let $V(\phi) = K \phi^{-\alpha}$, 
and we have re-introduced $c$ for clarity. We may break up equation~(\ref{eq:phi}) 
using standard operator splitting techniques:
\begin{eqnarray}
	\label{eq:splitPhi1}
		{\phi}^n \rightarrow {\phi}^{(1)} & : & \frac{\partial^2 \phi}{\partial t^2} - \frac{c^2}{a^2} \nabla^2 \phi = 0 \\		
	\label{eq:splitPhi2}
		{\phi}^{(1)} \rightarrow {\phi}^{n+1} & : & \frac{\partial^2 \phi}{\partial t^2} + \frac{d V}{d \phi} 
								+ 3 \frac{\dot{a}}{a} \frac{\partial \phi}{\partial t} + \frac{\rho}{\phi}a^{-3} = 0.
\end{eqnarray}

Here and below, superscripts are temporal indices and subscripts 
will denote the index of the location on the mesh.  
We solve Eq.(\ref{eq:splitPhi1}) by reducing it to a set 
of two 1st-order equations: 
\begin{eqnarray}
	\label{eq:splitWave1}
  	\frac{\partial \phi^{(1)}}{\partial t} & = & \dot{\phi}^{n} \\
  \label{eq:splitWave2}	
  	\frac{\partial \dot {\phi}^{(1)}}{\partial t} & = & - \frac{c^2}{a^2} \nabla^2 \phi^{n}.
\end{eqnarray}
 
Since the sound speed is so high ($v_s = c/a$), 
we found that a direct integration scheme was highly unstable 
except for unworkably small $\Delta t$. So, 
we use a midpoint method for Eq.(\ref{eq:splitWave1}) 
and a fully implicit method for equation~(\ref{eq:splitWave2}) :
\begin{eqnarray}
	\label{eq:midpoint}
  	\phi^{1}_{i} & = & \phi^{n}_i  + \frac{1}{2} \Delta t \left(  \dot{\phi}^{n}_i + \dot{\phi}^{(1)}_i \right)\\
  \label{eq:implicit}	
  	\dot{\phi}^{(1)}_i & = & \dot{\phi}^{n}_i
  			 + \frac{c^2 \Delta t}{\Delta x^2 \left( a^{(1)}\right)^2} 
  													\left( \phi^{(1)}_{i+1} - 2 \phi^{(1)}_{i} + \phi^{(1)}_{i-1} \right).
\end{eqnarray}

We substitute our expression for $\phi^{(1)}_{i}$ into equation~(\ref{eq:implicit}), yielding the matrix equation $Ax=b$, where
\begin{equation}
x_i \equiv \dot{\phi}_i^{(1)},
\end{equation}
\begin{eqnarray}
b_i \equiv \dot{\phi}^{n}_{i} & + & \sigma \phi^{n}_{i+1} + \gamma \sigma \dot{\phi}^{n}_{i+1} \nonumber \\
															& - & 2 \sigma \phi^{n}_{i} -2 \gamma \sigma \dot{\phi}^{n}_{i} \\
															& + & \sigma \phi^{n}_{i-1} + \gamma \sigma \dot{\phi}^{n}_{i-1} \nonumber,
\end{eqnarray}
and
\begin{eqnarray}
A = \left( \begin{array}{ccccc}
											1+2 \sigma \gamma & -\sigma \gamma & 0 & \cdots & -\sigma \gamma  \\
											-\sigma \gamma & 1+2 \sigma \gamma & -\sigma \gamma & \cdots & 0  \\
											0 & -\sigma \gamma & 1+2 \sigma \gamma & -\sigma \gamma & 0  \\
											\cdots & \cdots & \cdots & \cdots & \cdots  \\ 
											-\sigma \gamma & \cdots & 0 & -\sigma \gamma & 1+2 \sigma \gamma \end{array} 
														\right) \nonumber.
\end{eqnarray}
Note the periodic boundary conditions. We have defined
\begin{eqnarray}
\sigma & \equiv & \frac{c^2}{\left( a^{(1)}\right)^2} \frac{\Delta t}{\Delta x^2}, \\
\gamma & \equiv & \frac{1}{2} \Delta t. \nonumber
\end{eqnarray}
We solved this matrix-vector equation exactly using a Thomas algorithm~\citep{Thomas}, modified for 
periodic boundaries via the Sherman-Morrison formula~\citep{Sherman}.

We may now update our solution through equation~(\ref{eq:splitPhi2}) by expanding in a Taylor series:
\begin{equation}
\dot{\phi}^{n+1} = \dot{\phi}^{(1)} + \ddot{\phi}^{(1)} \Delta t 
									+ \frac{1}{2} \frac{d^3 \phi^{(1)}}{dt^3} \Delta t^2 + O(\Delta t^3),
\end{equation}
where
\begin{eqnarray*}
\ddot{\phi}^{(1)} & = & K \alpha \left( {\phi^{n}}\right)^{-\alpha-1} 
											- 3 \frac{\dot{a}^{n}}{a^{n}} \dot{\phi}^{(1)} 
											- \frac{\rho}{\phi^{(1)}}(a^{n})^{-3}
\end{eqnarray*}
and
\begin{eqnarray*}
\frac{d^3 \phi^{(1)}}{dt^3}& = & -K \alpha (\alpha+1) \left(\phi^{n} \right)^{-\alpha-2} \dot{\phi}^{(1)} 
											- 3 \frac{\ddot{a}^{n}}{a^{n}} \dot{\phi}^{(1)} \\
 									&\mbox{ }&			+ 12 \left( \frac{\dot{a}^{n}}{a^{n}}  \right)^2 \dot{\phi}^{(1)} 
													- \frac{\dot \rho}{\phi^{(1)}}(a^{n})^{-3} \\ 
									&\mbox{ }&			- \frac{\rho}{\left( \phi^{(1)} \right)^2}(a^{n})^{-3} \dot{\phi}^{(1)}
												+3 \frac{\rho}{\phi^{(1)}}(a^{n})^{-4} (\dot{a}^{n}).
\end{eqnarray*}
Here, $\dot \rho = \left( \rho^{n} - \rho^{n-1}\right)/dt $.

At any step $n$, after computing $\phi^{n}$ across the grid, 
we may update the scale factor using a Taylor series:
\begin{equation}
a^{n+1} = a^n + \dot{a}^n \Delta t + \frac{1}{2} \ddot{a}^n \Delta t^2 + O(\Delta t^3),
\end{equation}

where $\dot{a}$ and $\ddot{a}$ are given by the Friedmann equation and Friedmann energy equation, respectively:
\begin{eqnarray}
\label{eq:a}
\dot{a}^{n} & = & \left[ H_0^2 \Omega_m \frac{ \overline{\phi} }{ \overline{\phi}_0} a^{-1} 
									+ \frac{8 \pi G}{3} a^{2} \rho_\phi \right]^{1/2} \\
\ddot{a}^{n} & = & -\frac{1}{2} H_0^2 \Omega_m \frac{\overline{\phi} }{\overline{\phi}_0} a^{-2} - \frac{4 \pi G}{3} (1+3w)\rho_\phi a.
\end{eqnarray}

Ignoring fluctuations, the equation of state parameter, $w$, is 
\begin{equation}
w \equiv \frac{p_\phi}{\rho_\phi} = \frac{\frac{1}{2}  \overline{\dot \phi}^2 -  V(\overline{\phi})}
																		{\frac{1}{2} \overline{\dot \phi}^2 + V(\overline{\phi})}.
\end{equation}

\bibliographystyle{apj}	
\bibliography{ms}		
\nocite{*}

\end{document}